\documentclass[oneside,reqno]{amsart}
\usepackage{amsmath}
\usepackage{amssymb}
\usepackage{xspace}
\usepackage[mathscr]{eucal}

\newcommand{\mat}[3]{\ensuremath{
																								\left \langle  \vphantom{#2 #3}   #1   
                        \right|    					\, #2\,   
                        \left|    \vphantom{#2 #1} #3   
                        \right \rangle
                        								}
                     }
\newcommand{\bmat}[3]{\ensuremath{
																								\bigl \langle     #1   \bigr|    \, #2\,   \bigl|     #3   \bigr \rangle
                        									}
                     }

\newcommand{\ket}[1]{\ensuremath{		\left| #1 \right> 
																																			  }
																									}

\newcommand{\comm}[2]{\ensuremath{  \left[ #1, #2 \right] }}
\newcommand{\bcomm}[2]{\ensuremath{							\bigl[ #1 , #2 \bigr]							}}
\newcommand{\pt}{   \ensuremath{     \phi_{\mathrm{ap}}     }}
\newcommand{\xOp}{ \ensuremath{  \hat{x}  }}
\newcommand{\pOp}{ \ensuremath{  \hat{p}  }}
\newcommand{\xiOp}{ \ensuremath{  \hat{x}_{\mathrm{i}}  }\xspace}
\newcommand{\piOp}{ \ensuremath{  \hat{p}_{\mathrm{i}}  }\xspace}
\newcommand{\xfOp}{ \ensuremath{  \hat{x}_{\mathrm{f}}  }\xspace}
\newcommand{\pfOp}{ \ensuremath{  \hat{p}_{\mathrm{f}}  }\xspace}
\newcommand{\MxOp}{  \ensuremath{   \hat{\mu}_{\mathrm{X}}    } }
\newcommand{\MpOp}{  \ensuremath{   \hat{\mu}_{\mathrm{P}}    } }

\newcommand{\MpiOp}{  \ensuremath{   \hat{\mu}_{\mathrm{Pi}}    } }
\newcommand{\MxfOp}{  \ensuremath{   \hat{\mu}_{\mathrm{Xf}}    } }
\newcommand{\MpfOp}{  \ensuremath{   \hat{\mu}_{\mathrm{Pf}}    } }
\newcommand{\Mx}{   \ensuremath{   \mu_{\mathrm{X}}  } }
\newcommand{\Mp}{   \ensuremath{   \mu_{\mathrm{P}}  } }
\newcommand{\PxOp}{  \ensuremath{   \hat{\pi}_{\mathrm{X}}    } }

\newcommand{\PxfOp}{  \ensuremath{   \hat{\pi}_{\mathrm{Xf}}    } }

\newcommand{\Px}{   \ensuremath{   \pi_{\mathrm{X}}  } }

\newcommand{\yOp}[1]{  \ensuremath{ \hat{y}_{#1} } }
\newcommand{\yv}[1]{  \ensuremath{ y_{#1} } }

\newcommand{\Exi}{\ensuremath{       \hat{\epsilon}_{\mathrm{Xi}}       }\xspace}
\newcommand{\Epi}{\ensuremath{       \hat{\epsilon}_{\mathrm{Pi}}       }\xspace}

\newcommand{\Exf}{\ensuremath{       \hat{\epsilon}_{\mathrm{Xf}}       }\xspace}
\newcommand{\Epf}{\ensuremath{       \hat{\epsilon}_{\mathrm{Pf}}       }\xspace}

\newcommand{\Dx}{\ensuremath{       \hat{\delta}_{\mathrm{X}}       }\xspace}
\newcommand{\Dp}{\ensuremath{       \hat{\delta}_{\mathrm{P}}       }\xspace}
\newcommand{\RErr}{\ensuremath{     \Delta_{\mathrm{ei}}  }\xspace}
\newcommand{\PErr}{\ensuremath{     \Delta_{\mathrm{ef}}  }\xspace}
\newcommand{\Dist}{\ensuremath{     \Delta_{\mathrm{d}}  }\xspace}
\begin{document}
\begin{titlepage}
\begin{center}
\bfseries
THE ERROR PRINCIPLE
\end{center}
\vspace{1 cm}
\begin{center}
D M APPLEBY 
\end{center}
\begin{center}
Department of Physics, Queen Mary and
		Westfield College,  Mile End Rd, London E1 4NS, UK
\end{center}
\vspace{0.3 cm}
\begin{center}
(e-mail:  D.M.Appleby@qmw.ac.uk) 
\end{center}
\vspace{1.5 cm}
\begin{center}
\textbf{Abstract}\\
\vspace{0.35 cm}
\parbox{10.5 cm }{   The problem of characterising the accuracy of, and disturbance caused by a joint measurement 
                     of position and momentum is investigated.  In a previous paper the problem was discussed in
                     the context of the unbiased measurements considered by Arthurs and Kelly.  
                     It is now shown, that suitably modified versions of these results hold for
                     a much larger class of simultaneous measurements. 
                     The approach is a development of that adopted by Braginsky 
                     and Khalili in the case of a single measurement of position only.  
                     A distinction is made between the errors of retrodiction and the errors of prediction.   
                     Two error-error relationships and four error-disturbance relationships are derived, 
                     supplementing the Uncertainty Principle usually so-called. In the general case it is necessary to 
                     take into account the range of the measuring apparatus.  Both the ideal case, of an instrument 
                     having infinite range, and the case of a real instrument, for which the range is finite, 
                     are discussed.  
                 }\\     
\vspace{1.0 cm}
\parbox{10.5 cm }{
\textbf{PACS number:}  03.65.Bz}
\end{center}
\vspace{2 cm}
\begin{center}
Report no. QMW-PH-98-13
\end{center}
\end{titlepage}
\section{Introduction}
\label{sec:  intro}
Heisenberg's~\cite{Heis1} formulation of the Uncertainty Principle was one of the key steps
in the development of Quantum Mechanics.
Nevertheless, seventy years after the publication
of his original paper, there remain a number of obscurities regarding its 
interpretation~\cite{Hilgevoord}.

In contemporary discussions the Uncertainty Principle is usually identified with the
statement
\begin{equation}
    \Delta x \Delta p \ge \frac{\hbar}{2}
\label{eq:  UncertPrinc}
\end{equation}
where $\Delta x$, $\Delta p$ are the standard deviations
\begin{equation}
\begin{split}
    \Delta x & = \left( \mat{\psi}{\xOp^2}{\psi} - \mat{\psi}{\xOp}{\psi}^2 \right)^{\frac{1}{2}} \\
    \Delta p & = \left( \mat{\psi}{\pOp^2}{\psi} - \mat{\psi}{\pOp}{\psi}^2 \right)^{\frac{1}{2}}
\end{split}
\label{eq:  UncertDef}
\end{equation}
In his original paper Heisenberg suggested that the quantities
$\Delta x$, $\Delta p$ appearing in Eq.~(\ref{eq:  UncertPrinc}) may be interpreted 
as experimental errors, and that the Uncertainty Principle
represents a fundamental constraint on the accuracy achievable in a simultaneous measurement
of position and momentum.  At least, that is what he has often been taken to 
have suggested (Heisenberg's own phraseology is somewhat ambiguous).  
In the words of Bohm~\cite{Bohm}:
\begin{quote}
  If a measurement of position is made with accuracy $\Delta x$, and if a measurement 
  of momentum is made \emph{simultaneously} with accuracy $\Delta p$, then the product of the 
  two errors can never be smaller than a number of order $\hbar$.
\end{quote}
Is this is a legitimate interpretation of Eq.~(\ref{eq:  UncertPrinc})?
The question has been discussed by 
Ballentine~\cite{Ballentine}, Prugove\v{c}ki~\cite{Prugovecki}, 
Busch~\cite{BuschUncert},
W\'{o}dkiewicz~\cite{Wodkiewicz}, Hilgevoord and 
Uffink~\cite{Hilgevoord}, Raymer~\cite{Raymer} and de Muynck \emph{et al}~\cite{deMuynck}.  
The consensus seems to be, that the quantities $\Delta x$, $\Delta p$ defined in
Eq.~(\ref{eq:  UncertDef}) cannot be regarded as experimental errors because they
are intrinsic properties of the isolated system.  An experimental error, by contrast,
should depend, not only on the state of the system, but also on the state of the 
apparatus, and the nature of the measurement interaction.  
Hilgevoord and Uffink~\cite{Hilgevoord} have further remarked, that in Heisenberg's microscope  
argument, it is only the position of the particle which is 
measured.  Although it is true that Heisenberg alludes to the possibility of performing 
simultaneous measurements of position \emph{and} momentum, such measurements form no part 
of his actual argument.

It follows from all this, that the statement of Bohm's just quoted cannot be identified 
with the Uncertainty Principle usually so-called.  Rather, it represents (if true)
an independent physical principle:  the Error Principle, as it might be called.

The problem we now face is, that although the Error Principle as stated 
by Bohm is intuitively
quite plausible, it cannot be regarded as rigorously established.  In order to 
establish it two things are necessary.  In the first place, we need 
to define precisely what is meant by the accuracy of a simultaneous measurement
process.  In the second place, we need to derive a bound on the accuracy,
starting from the fundamental principles
of Quantum Mechanics.
The problem is of some interest, in view of the importance that simultaneous
measurements now have in the field of quantum 
optics~\cite{Arthurs,ArtGood,MoreArtKel,SimMeas,LeonRev,LeonhardtBook,BuschBook}.

An approach to the problem which has attracted a good deal of attention over the years
is the one based on positive operator valued measures and the concept of a 
``fuzzy'' or ``stochastic'' phase 
space~\cite{Prugovecki,BuschUncert,deMuynck,Davies,Holevo,BuschRetPre,ErrPrinciple,Ban}.
For a recent review see Busch \emph{et al}~\cite{BuschBook}.
This approach has recently been criticised by Uffink~\cite{Uffink}.

In a previous paper~\cite{Appleby} we adopted a rather different approach.  We began with
Braginsky and Khalili's~\cite{Braginsky} analysis of single measurements of
$x$ or $p$ by themselves, and extended it to a  class of simultaneous measurement
processes:  namely, the class of unbiased measurement processes, for which 
the systematic errors are all zero.
Our analysis depended on making a  distinction
between the retrodictive and predictive (or determinative and preparative) aspects of a 
measurement~\cite{Hilgevoord,BuschRetPre,RetroPred}.  
We  accordingly defined two different kinds
of error:  the errors of retrodiction, $\RErr x$ and $\RErr p$, describing the
accuracy with which the result of the measurement reflects the initial state
of the system; and the errors of prediction, $\PErr x$ and $\PErr p$, describing
the accuracy with which the result of the measurement reflects the final
state of the system.  Corresponding to these two kinds of error we  derived 
two inequalities: a
retrodictive error relationship
\begin{equation}
    \RErr x  \, \RErr p  \ge \frac{\hbar}{2}
\label{eq:  RetErrRel}
\end{equation}
and a predictive error relationship
\begin{equation}
    \PErr x  \, \PErr p  \ge \frac{\hbar}{2}
\label{eq:  PreErrRel}
\end{equation}
Eqs.~(\ref{eq:  RetErrRel}) and~(\ref{eq:  PreErrRel}) jointly comprise a precise
statement of the semi-intuitive Error Principle discussed above.

Following Braginsky and Khalili we also defined two quantities $\Dist x$, $\Dist p$
describing the disturbance of the system by the measurement; and we
derived the four error-disturbance relationships
\begin{equation}
\begin{aligned}
  \RErr x \, \Dist p  & \ge \frac{\hbar}{2} & \hspace{0.5 in} \PErr x \, \Dist p & \ge \frac{\hbar}{2} \\
  \RErr p \, \Dist x  & \ge \frac{\hbar}{2} & \hspace{0.5 in} \PErr p \, \Dist x & \ge \frac{\hbar}{2}
\end{aligned}
\label{eq:  ErrDistRels}
\end{equation}
These relationships provide a precise statement of the principle, that a decrease in the error
of the measurement of one observable can only be achieved at the cost of a corresponding increase
in the disturbance of the canonically conjugate observable.

The relationships above, together with Eq.~(\ref{eq:  UncertPrinc}), comprise a total
of seven inequalities, all of which are needed if one wants to capture the full intuitive content
of Heisenberg's original paper~\cite{Heis1}.

Arthurs and Kelly~\cite{Arthurs} have shown, that in the case of a retrodictively unbiased
joint measurement process (\emph{i.e.}\ a process for which the final state expectation values of the 
pointer positions coincide with the initial state expectation values of the position 
and momentum), one has
\begin{equation}
  \Delta \Mx \, \Delta \Mp  \ge \hbar
\label{eq:  ArtKelRel}
\end{equation}
where the quantities on the right hand side are the final state uncertainties for the pointer
positions $\Mx$ and $\Mp$ (also see Arthurs and Goodman~\cite{ArtGood}, 
W\'{o}dkiewicz~\cite{Wodkiewicz}, Raymer~\cite{Raymer} and 
Leonhardt and Paul~\cite{LeonOp}).  In ref.~\cite{Appleby} we showed that the
Arthurs-Kelly relationship can be deduced from the retrodictive error relationship.

The unsatisfactory feature of the arguments given in ref.~\cite{Appleby} is
that they only serve to establish the above inequalities 
for a limited class of measurement processes.  That is,
we only proved Eq.~(\ref{eq:  RetErrRel}) on the assumption that the measurement
is retrodictively unbiased, and Eq.~(\ref{eq:  ErrDistRels}) on the still more
restrictive assumption that the measurement is both retrodictively and predictively
unbiased.  Our purpose in the following is to show, that with a suitable 
modification of the definitions, these relationships continue to hold
for a very much larger class of measurement processes.
\section{Simultaneous Measurement Processes}
\label{sec:  SimMeasProcs}
We begin by characterising the class of measurement processes which we are going to discuss.

Consider a system, with state space $\mathscr{H}_{\mathrm{sy}}$, interacting with an apparatus,
with state space $\mathscr{H}_{\mathrm{ap}}$.
The system is assumed to have one degree of freedom, with position $\xOp$ and momentum $\pOp$,
satisfying the commutation relationship
\begin{equation}
  \comm{\xOp}{\pOp} = i \hbar
\label{eq:  sysCCRs}
\end{equation}
The apparatus is assumed to be characterised by two pointer observables $\MxOp$ (measuring the position
of the system) and $\MpOp$ (measuring the momentum of the system), together with $n$ 
other observables $\yOp{1},\yOp{2}, \dots \yOp{n}$.  These $n+2$ operators constitute
a complete set of commuting observables describing the state of the apparatus.  They also
commute with the system observables $\xOp$, $\pOp$.

It is assumed that the system$+$apparatus is initially in a product state of the form
$\ket{\psi \otimes \pt}$, where
$\ket{\psi} \in \mathscr{H}_{\mathrm{sy}}$ is the initial state of the system 
and $\ket{\pt} \in \mathscr{H}_{\mathrm{ap}}$ is the intial state of the apparatus.
The unitary evolution operator describing the measurement interaction will be denoted 
$\hat{U}$.  The final state of the system$+$apparatus is 
$\hat{U} \ket{\psi \otimes \pt}$. The  probability distribution of the measured values is 
\begin{equation*}
  \rho \left( \Mx, \Mp \right)
= \int dx d \yv{1} \dots d\yv{n} \, \bigl|\bmat{x,\Mx,\Mp,\yv{1},\dots,\yv{n}}{\hat{U}}{\psi\otimes\pt}\bigr|^2
\end{equation*}

In ref.~\cite{Appleby} we assumed that the measurement process
was unbiased, so that
\begin{align*}
    \bmat{\psi \otimes \pt}{\hat{U}^{\dagger} \MxOp \hat{U} }{\psi \otimes \pt}
=   \bmat{\psi \otimes \pt}{\hat{U}^{\dagger} \xOp \hat{U} }{\psi \otimes \pt}
=   \bmat{\psi \otimes \pt}{\xOp }{\psi \otimes \pt}
\\
\intertext{and}
    \bmat{\psi \otimes \pt}{\hat{U}^{\dagger} \MpOp \hat{U} }{\psi \otimes \pt}
=   \bmat{\psi \otimes \pt}{\hat{U}^{\dagger} \pOp \hat{U} }{\psi \otimes \pt}
=   \bmat{\psi \otimes \pt}{\pOp }{\psi \otimes \pt}
\end{align*}
We make no such assumption here.

It may also be worth noting that we do not assume the existence of momenta canonically
conjugate to the pointer observables (as is the case in the Arthurs-Kelly process~\cite{Arthurs,MoreArtKel,LeonhardtBook},
for example).  In particular, we make no assumptions regarding 
the spectra of the pointer observables.
\section{Definition of the Errors and Disturbances}
\label{sec:  ErrDistDefs}
Let $\mathscr{O}$ be any of the Schr\"{o}dinger picture operators
$\xOp$, $\pOp$, $\MxOp$, $\MpOp$.  Let $\mathscr{O}_{\mathrm{i}} =\mathscr{O}$
be the corresponding Heisenberg picture operator at the instant the measurement interaction
begins; and let $\mathscr{O}_{\mathrm{f}} =\hat{U}^{\dagger}\mathscr{O}\hat{U}$ be 
the Heisenberg picture operator at the instant the interaction finishes.  Define the 
retrodictive error operators
\begin{equation}
    \Exi = \MxfOp - \xiOp \hspace{0.75 in} \Epi = \MpfOp - \piOp
\label{eq:  RetErrDef}
\end{equation}
the predictive error operators
\begin{equation}
    \Exf = \MxfOp - \xfOp \hspace{0.75 in} \Epf = \MpfOp - \pfOp
\label{eq:  PreErrDef}
\end{equation}
and the disturbance operators
\begin{equation}
    \Dx = \xfOp - \xiOp \hspace{0.75 in} \Dp = \pfOp - \piOp
\label{eq:  DistDef}
\end{equation}
Let $\mathscr{S}$ be the unit sphere in the system state space
$\mathscr{H}_{\mathrm{sy}}$.  We then define the maximal rms errors
of retrodiction
\begin{equation}
\begin{split}
      \RErr x 
  & = \sup_{\ket{\psi} \in \mathscr{S}} \Bigl( \bmat{\psi \otimes \pt}{\Exi^2}{\psi \otimes \pt}\Bigr)^{\frac{1}{2}} \\
      \RErr p 
  & = \sup_{\ket{\psi} \in \mathscr{S}} \Bigl( \bmat{\psi \otimes \pt}{\Epi^2}{\psi \otimes \pt}\Bigr)^{\frac{1}{2}}
\end{split}
\label{eq:  MaxRMSRetErr}
\end{equation}
the maximal rms errors of prediction
\begin{equation}
\begin{split}
      \PErr x 
  & = \sup_{\ket{\psi} \in \mathscr{S}} \Bigl( \bmat{\psi \otimes \pt}{\Exf^2}{\psi \otimes \pt}\Bigr)^{\frac{1}{2}} \\
      \PErr p 
  & = \sup_{\ket{\psi} \in \mathscr{S}} \Bigl( \bmat{\psi \otimes \pt}{\Epf^2}{\psi \otimes \pt}\Bigr)^{\frac{1}{2}}
\end{split}
\label{eq:  MaxRMSPreErr}
\end{equation}
and the maximal rms disturbances
\begin{equation}
\begin{split}
      \Dist x 
  & = \sup_{\ket{\psi} \in \mathscr{S}} \Bigl( \bmat{\psi \otimes \pt}{\Dx^2}{\psi \otimes \pt}\Bigr)^{\frac{1}{2}} \\
      \Dist p 
  & = \sup_{\ket{\psi} \in \mathscr{S}} \Bigl( \bmat{\psi \otimes \pt}{\Dp^2}{\psi \otimes \pt}\Bigr)^{\frac{1}{2}}
\end{split}
\label{eq:  MaxRMSDist}
\end{equation}
We discussed the physical interpretation of these quantities in 
ref.~\cite{Appleby}.  The reader may confirm that this interpretation 
continues to be valid in the present more general context.

It should be noted, that in these definitions, the supremum is only taken over all
normalised initial system states.  The initial apparatus state is held fixed.  
The quantities $\RErr x$, $\RErr p$, $\PErr x$, $\PErr p$, $\Dist x$, $\Dist p$
are therefore functions of the initial apparatus state.

It should also be noted that the  definitions just given differ slightly from those
in ref.~\cite{Appleby}, in that we did not
previously take the supremum over all initial system states.
Some such change in the definitions is essential, if the error-error and
error-disturbance relationships proved in 
ref.~\cite{Appleby}, for the special case of unbiased measurement 
processes, are to be generalised, so as to apply to the larger
class of processes considered in this paper.
As we show in the
appendix, if one drops the requirement that the measurement be unbiased,
then it is possible to find processes such that, with a suitable choice of initial 
state $\ket{\psi}$,
\begin{equation*}
    \bmat{\psi \otimes \pt}{\Exi^2}{\psi \otimes \pt} 
    \bmat{\psi \otimes \pt}{\Epi^2}{\psi \otimes \pt} 
= 0
\end{equation*}
It is only when one takes the supremum over all states $\ket{\psi}$ 
that one gets the inequalities
of Eqs.~(\ref{eq:  RetErrRelB}) and~(\ref{eq:  DistErrRelB}).

As discussed in reference~\cite{Appleby}, the quantity 
$\left(\bmat{\psi \otimes \pt}{\Exi^2}{\psi \otimes \pt}\right)^{\frac{1}{2}}$
represents the rms retrodictive error in the measurement of $x$ when the system
is initially in the state $\ket{\psi}$.  The quantity
$\RErr x$ (as defined above) consequently represents the maximum rms
error obtained, when the system is allowed to range over every possible
initial state.  Similarly with the
quantities $\RErr p$, $\PErr x$, $\PErr p$, $\Dist x$, $\Dist p$.

It is easy to think of measurement interactions for which the errors and disturbances
defined in Eqs.~(\ref{eq:  MaxRMSRetErr}--\ref{eq:  MaxRMSDist}) are  
finite (with an appropriate choice 
of initial apparatus state).  An example
of such a process is the Arthurs-Kelly process~\cite{Arthurs,MoreArtKel,LeonhardtBook} 
(see ref.~\cite{self2}).  In the case of the Arthurs-Kelly process 
$\bmat{\psi \otimes \pt}{\mathscr{O}^2}{\psi\otimes \pt}$,
with $\mathscr{O}$ any error or disturbance operator, is independent
of the state $\ket{\psi}$---which means, that in the particular case
of the Arthurs-Kelly process, the definitions of 
$\RErr x$, $\RErr p$, $\PErr x$, $\PErr p$, $\Dist x$, $\Dist p$ which are employed in 
this paper coincide with the definitions used in 
refs.~\cite{Appleby,self2}).  It should, however, be observed that interactions for which
this is true are somewhat idealised.  A real measuring instrument will have a finite range.  If
the initial system state expectation values
$\mat{\psi}{\xiOp}{\psi}$ and $\mat{\psi}{\piOp}{\psi}$ are a long way outside the
range of the instrument, then the errors and disturbances may not be small.
Consequently, in the case
of a real measuring instrument, the quantities
defined in 
Eqs.~(\ref{eq:  MaxRMSRetErr}--\ref{eq:  MaxRMSDist}) may well be infinite, or at 
least very large.  To put it another way, in the case of a real measuring instrument, 
these quantities 
do not correspond very closely
to one's intuitive idea of the accuracy of and disturbance caused by
a realistic measurement process.  In section~\ref{sec:  FinRange} we show how 
the definitions can be modified, so as to obviate this difficulty.
\section{Commutators}
\label{sec:  CommRels}
We have, as an immediate consequence of the definitions,
\begin{equation}
    \comm{\Exf}{\Epf} = i \hbar
\label{eq:  PreErrComRel}
\end{equation}
The other commutators between the error and disturbance operators give more difficulty.  
This is because the  retrodictive error and disturbance operators mix
Heisenberg picture observables defined at different times.  It turns out, however, that it is 
possible to express every remaining commutator of interest 
in terms of commutators between one of the
operators $\Exi$, $\Epi$, $\Exf$, $\Epf$, $\Dx$, $\Dp$
and one of the operators $\xiOp$, $\piOp$.  The significance of this result is that 
$\xiOp$, $\piOp$ generate translations in the system phase space.

In fact
\begin{align}
    \bcomm{\Exi}{\Epi} & = \bcomm{\left(\MxfOp - \xiOp\right)}{\left(\MpfOp - \piOp\right)} \notag\\
                      & = i \hbar - \bcomm{\xiOp}{\MpfOp} + \bcomm{\piOp}{\MxfOp} \notag \\
                      & = i \hbar - \bcomm{\xiOp}{\left(\piOp + \Epi\right)} + \bcomm{\piOp}{\left(\xiOp + \Exi\right)}\notag\\
                      & = - i \hbar - \bcomm{\xiOp}{\Epi} + \bcomm{\piOp}{\Exi} \label{eq:  RetErrComRel}
\end{align}
Similarly
{\allowdisplaybreaks
\begin{align}
\begin{split}
    & \bcomm{\Exi}{\Dp}  = - i \hbar - \bcomm{\xiOp}{\Dp} + \bcomm{\piOp}{\Exi} \\
    & \bcomm{\Dx}{\Epi}  = - i \hbar - \bcomm{\xiOp}{\Epi} + \bcomm{\piOp}{\Dx} \\
\end{split}
\label{eq:  ErrDisComRelA}
\\
\intertext{and}
\begin{split}
    & \bcomm{\Exf}{\Dp}  = - i \hbar + \bcomm{\piOp}{\Exf} \\
    & \bcomm{\Dx}{\Epf}  = - i \hbar - \bcomm{\xiOp}{\Epf}
\end{split}
\label{eq:  ErrDisComRelB}
\end{align}
}
\section{Error and Error-Disturbance Relationships}
\label{sec:  ErrAndErrDistRels}
We have, as an immediate consequence of Eq.~(\ref{eq:  PreErrComRel}),
\begin{equation}
   \PErr x \, \PErr p \ge \frac{\hbar}{2}
\label{eq:  PreErrRelB}
\end{equation}
For the remaining relationships we have to work a little harder.  Let $\ket{\psi}$
be any normalised state $\in \mathscr{H}_{\mathrm{sy}}$.  Let 
\begin{equation*}
    \hat{D}_{xp} = \exp \left[ \tfrac{i}{\hbar} \left( p \xOp - x \pOp \right) \right]
\end{equation*}
be the system  phase space displacement operator, and define
\begin{equation*}
   \ket{\psi_{xp}} = \hat{D}_{xp} \ket{\psi}
\end{equation*}
We have
\begin{equation*}
\begin{aligned}
      i \hbar \frac{\partial}{\partial x} \hat{D}_{xp}^{\vphantom{\dagger}} 
  & = \left( \pOp - \tfrac{1}{2} p\right) \hat{D}_{xp}^{\vphantom{\dagger}} & \hspace{0.75 in}
      - i \hbar \frac{\partial}{\partial x} \hat{D}_{xp}^{\dagger}
  & = \hat{D}_{xp}^{\dagger} \left( \pOp - \tfrac{1}{2} p\right) \\
      - i \hbar \frac{\partial}{\partial p} \hat{D}_{xp}^{\vphantom{\dagger}} 
  & = \left( \xOp - \tfrac{1}{2} x\right) \hat{D}_{xp}^{\vphantom{\dagger}} & \hspace{0.75 in}
      i \hbar \frac{\partial}{\partial p} \hat{D}_{xp}^{\dagger}
  & = \hat{D}_{xp}^{\dagger} \left( \xOp - \tfrac{1}{2} x\right) 
\end{aligned}
\end{equation*}
In view of Eq.~(\ref{eq:  RetErrComRel}) we then have
\begin{equation}
  \bmat{\psi_{xp} \otimes \pt}{\comm{\Exi}{\Epi}}{\psi_{xp} \otimes \pt}
= - i \hbar \left( 1 + \boldsymbol{\nabla}  \boldsymbol{\cdot} \mathbf{v}\right)
\label{eq:  RetErrComRelB}
\end{equation}
where $\mathbf{v}$ is the vector
\begin{equation*}
   \mathbf{v} = \begin{pmatrix} 
                        \bmat{\psi_{xp} \otimes \pt}{\Exi}{\psi_{xp} \otimes \pt} \\
                        \bmat{\psi_{xp} \otimes \pt}{\Epi}{\psi_{xp} \otimes \pt}
                \end{pmatrix}
\end{equation*}
and $\boldsymbol{\nabla}$ is the phase space gradient operator
\begin{equation*}
  \boldsymbol{\nabla} = \begin{pmatrix}
                         \frac{\partial}{\partial x} \\ \frac{\partial}{\partial p} 
                         \end{pmatrix}
\end{equation*}
Now consider the box-shaped region $\mathscr{R}$ in phase space, 
with vertices at $\left(\frac{L}{2},\frac{P}{2}\right)$, $\left(-\frac{L}{2},\frac{P}{2}\right)$,
$\left(-\frac{L}{2},-\frac{P}{2}\right)$, $\left(\frac{L}{2},-\frac{P}{2}\right)$.  Let
$\mathscr{C}$ be its boundary.  We have
{\allowdisplaybreaks
\begin{align}
    \RErr x \, \RErr p 
  & \ge \frac{1}{2 L P} 
    \int_{\mathscr{R}} dx dp \, 
            \left| \bmat{\psi_{xp} \otimes \pt}{\comm{\Exi}{\Epi}}{\psi_{xp} \otimes \pt} \right|  
\notag \\
  & \ge \frac{1}{2 L P}
      \left| \int_{\mathscr{R}} dx dp \, 
                \bmat{\psi_{xp} \otimes \pt}{\comm{\Exi}{\Epi}}{\psi_{xp} \otimes \pt}
      \right|  
\notag \\
  & \ge  \frac{\hbar}{2} 
         \left( 1 - \frac{1}{LP} \left| \int_{\mathscr{R}} dx dp \, \boldsymbol{\nabla \cdot} \mathbf{v} \right| \right)  
\notag \\
  & =  \frac{\hbar}{2}
         \left( 1 - \frac{1}{LP} \left| \int_{\mathscr{C}} d s \, \mathbf{n} \boldsymbol{\cdot} \mathbf{v} \right| \right) 
\notag \\
  & \ge \frac{\hbar}{2}
         \left( 1 - \frac{2}{L} \RErr x - \frac{2}{P} \RErr p \right)
\label{eq:  RetErrDerive}
\end{align}
where $ds$ is the line element and $\mathbf{n}$ is the outward-pointing unit normal along $\mathscr{C}$.
Taking the limit as $L,P \rightarrow \infty$ we deduce
}    
\begin{equation}
  \RErr x \, \RErr p \ge \frac{\hbar}{2}
\label{eq:  RetErrRelB}
\end{equation}
whenever the left hand side is defined (\emph{i.e.} whenever it is not of the form $0\times \infty$).

Starting from Eqs.~(\ref{eq:  ErrDisComRelA}) and~(\ref{eq:  ErrDisComRelB}) 
we deduce, by essentially the same argument,
\begin{equation}
\begin{aligned}
  \RErr x \, \Dist p & \ge \frac{\hbar}{2}  & \hspace{0.75 in} \PErr x \, \Dist p & \ge \frac{\hbar}{2} \\
  \RErr p \, \Dist x & \ge \frac{\hbar}{2}  & \hspace{0.75 in} \PErr p \, \Dist x & \ge \frac{\hbar}{2}
\end{aligned}
\label{eq:  DistErrRelB}
\end{equation}
whenever the products are defined.

It should be noted, that although the relationships proved in this section have
the same form as the corresponding relationships proved in
ref.~\cite{Appleby}, they do not have  the same content, since
the quantities
$\RErr x$, $\RErr p$, $\PErr x$, $\PErr p$, $\Dist x$, $\Dist p$ appearing
in them are not defined in the same way (in ref.~\cite{Appleby} we did
not take a supremum over all normalised initial system states when defining
the errors and disturbances.  See 
Section~\ref{sec:  ErrDistDefs} above, and Section~\ref{sec:  Unbiased} immediately following).

In the Introduction we remarked, that in the case of the retrodictively unbiased
measurement processes considered in ref.~\cite{Appleby},
the Arthurs-Kelly relationship [Eq.~(\ref{eq:  ArtKelRel}) above]
is a consequence of the retrodictive error relationship.  It is an interesting
question, which we have not as yet been able to resolve, whether it is possible
to deduce an Arthurs-Kelly type bound from Eq.~(\ref{eq:  RetErrRelB}),
applying to the much more general class of measurement processes considered
in this paper.
\section{Unbiased Measurements}
\label{sec:  Unbiased}
Suppose that the measurement process is retrodictively unbiased, in the sense that 
\begin{equation*}
    \bmat{\psi \otimes \pt}{\Exi}{\psi \otimes \pt} 
=   \bmat{\psi \otimes \pt}{\Epi}{\psi \otimes \pt}
=   0
\end{equation*}
uniformly, for all $\ket{\psi} \in \mathscr{H}_{\mathrm{sy}}$ (but fixed $\ket{\pt}$).  Then the vector 
$\mathbf{v}$
appearing on the right hand side of Eq.~(\ref{eq:  RetErrComRelB}) is identically zero, and we have
\begin{equation*}
    \bmat{\psi \otimes \pt}{\Exi^2}{\psi \otimes \pt} \, \bmat{\psi \otimes \pt}{\Epi^2}{\psi \otimes \pt}
\ge   \frac{\hbar^2}{4}
\end{equation*}
uniformly, for all $\ket{\psi} \in \mathscr{H}_{\mathrm{sy}}$.

Suppose, in addition, that the measurement is predictively unbiased:
\begin{equation*}
    \bmat{\psi \otimes \pt}{\Exf}{\psi \otimes \pt} 
=   \bmat{\psi \otimes \pt}{\Epf}{\psi \otimes \pt}
=   0
\end{equation*}
for all $\ket{\psi}$.  Then we have, by a similar argument,
{\allowdisplaybreaks
\begin{align*}
        \bmat{\psi \otimes \pt}{\Exi^2}{\psi \otimes \pt} \, \bmat{\psi \otimes \pt}{\Dp^2}{\psi \otimes \pt}
& \ge   \frac{\hbar^2}{4} \\
        \bmat{\psi \otimes \pt}{\Exf^2}{\psi \otimes \pt} \, \bmat{\psi \otimes \pt}{\Dp^2}{\psi \otimes \pt}
& \ge   \frac{\hbar^2}{4} \\
        \bmat{\psi \otimes \pt}{\Epi^2}{\psi \otimes \pt} \, \bmat{\psi \otimes \pt}{\Dx^2}{\psi \otimes \pt}
& \ge   \frac{\hbar^2}{4} \\
        \bmat{\psi \otimes \pt}{\Epf^2}{\psi \otimes \pt} \, \bmat{\psi \otimes \pt}{\Dx^2}{\psi \otimes \pt}
& \ge   \frac{\hbar^2}{4} \\
\end{align*}
uniformly, for all $\ket{\psi}$.
}                

These are the results which we proved in ref.~\cite{Appleby} by a different method.
\section{Measurements with a Finite Range}
\label{sec:  FinRange}
Real measuring instruments are only designed to be used for a limited set
of initial system states.  For such an instrument one
expects the maximal rms errors and disturbances defined
in Eqs.~(\ref{eq:  MaxRMSRetErr}--\ref{eq:  MaxRMSDist}) to be infinite, or at least
very large.  This is because the supremum is taken over every possible initial system state, including 
those states for which the expected values of $\xOp$ and $\pOp$ are 
far outside the range of 
the instrument.  It follows that the quantities defined in Eqs.~(\ref{eq:  MaxRMSRetErr}--\ref{eq:  MaxRMSDist})
are poor indicators of the accuracies and disturbances to be expected when the instrument
is used in the manner in which it was designed to be used. 
In the case of a real measuring instrument, what interests us are the
maximum errors and disturbances obtained 
for a \emph{limited class} of initial system states---namely, the
class on which the instrument was designed to make measurements.  In this section we discuss an alternative
definition of the errors and disturbances which is more appropriate to such a case.

Suppose that the instrument is  designed to be accurate for initial system states $\ket{\psi}$ 
such that
\begin{align*}
    x_{0} - \tfrac{1}{2}L \le & \mat{\psi}{\xOp}{\psi} \le x_{0} + \tfrac{1}{2}L
    & \hspace{0.5 in}
    p_{0} - \tfrac{1}{2}P \le & \mat{\psi}{\pOp}{\psi} \le p_{0} + \tfrac{1}{2}P
\\
\intertext{and}
   & \Delta x  \le \sigma  & \hspace{0.5 in} & \Delta p  \le \tau
\end{align*}
for fixed constants $x_{0}$, $p_{0}$, $L$, $P$, $\sigma$, $\tau$ such that
$\sigma \tau \ge \frac{\hbar}{2}$.  Let $\mathscr{S}'$ be the set of
normalised states $\in \mathscr{H}_{\mathrm{sy}}$ which satisfy these conditions.  The errors and
disturbances appropriate for the description of this instrument are obtained by
taking the supremum over all normalised states $\ket{\psi} \in \mathscr{S}'$:
{\allowdisplaybreaks
\begin{align}
\begin{split}
     \RErr' x 
&  = \sup_{\ket{\psi} \in\mathscr{S}'} \Bigl(\bmat{\psi \otimes \pt}{\Exi^2}{\psi \otimes \pt}\Bigr)^{\frac{1}{2}} \\
     \RErr' p 
&  = \sup_{\ket{\psi} \in\mathscr{S}'} \Bigl(\bmat{\psi \otimes \pt}{\Epi^2}{\psi \otimes \pt}\Bigr)^{\frac{1}{2}} 
\end{split}
\label{eq:  FinRangeRetErrs}
\\
\begin{split}
     \PErr' x 
&  = \sup_{\ket{\psi} \in\mathscr{S}'} \Bigl(\bmat{\psi \otimes \pt}{\Exf^2}{\psi \otimes \pt}\Bigr)^{\frac{1}{2}} \\
     \PErr' p 
&  = \sup_{\ket{\psi} \in\mathscr{S}'} \Bigl(\bmat{\psi \otimes \pt}{\Epf^2}{\psi \otimes \pt}\Bigr)^{\frac{1}{2}} 
\end{split}
\label{eq:  FinRangePreErrs}
\\
\begin{split}
     \Dist' x 
&  = \sup_{\ket{\psi} \in\mathscr{S}'} \Bigl(\bmat{\psi \otimes \pt}{\Dx^2}{\psi \otimes \pt}\Bigr)^{\frac{1}{2}} \\
     \Dist' p 
&  = \sup_{\ket{\psi} \in\mathscr{S}'} \Bigl(\bmat{\psi \otimes \pt}{\Dp^2}{\psi \otimes \pt}\Bigr)^{\frac{1}{2}} 
\end{split}
\label{eq:  FinRangeDists}
\end{align}
It follows from Eq.~(\ref{eq:  PreErrComRel})
} 
\begin{equation*}
  \PErr' x \, \PErr' p \ge \frac{\hbar}{2}
\end{equation*}
Turning to the retrodictive error relationship, let $\ket{\psi}$ be any normalised state $\in \mathscr{H}_{\mathrm{sy}}$
such that
\begin{align*}
    \mat{\psi}{\xOp}{\psi} & = x_{0} & \hspace{0.75 in} \mat{\psi}{\pOp}{\psi}& = p_{0} \\ \intertext{and}
    \Delta x & \le \sigma & \hspace{0.75 in} \Delta p & \le \tau
\end{align*}
Let $\mathscr{R}$ be the box-shaped region of phase space 
with vertices $\left( x_{0} + \frac{L}{2},p_{0} + \frac{P}{2}\right)$, 
$\left( x_{0} - \frac{L}{2},p_{0} + \frac{P}{2}\right)$, $\left( x_{0} - \frac{L}{2},p_{0} - \frac{P}{2}\right)$, 
$\left( x_{0} + \frac{L}{2},p_{0} - \frac{P}{2}\right)$.  Then $\ket{\psi_{xp}} \in \mathscr{S}'$ for all
$(x,p) \in \mathscr{R}$.  We can now use an argument analogous to the one leading to 
Eq.~(\ref{eq:  RetErrDerive}) to deduce
\begin{equation*}
    \RErr' x \, \RErr' p \ge \frac{\hbar}{2} \left( 1 - \frac{2}{L} \RErr' x - \frac{2}{P} \RErr' p \right)
\end{equation*}
which can alternatively be written
\begin{equation}
    \left( \RErr' x + \frac{\hbar}{P} \right) \, \left( \RErr' p + \frac{\hbar}{L} \right)
\ge \frac{\hbar}{2} \left( 1 + \frac{2 \hbar}{L P} \right)
\label{eq:  RetErrRelC}
\end{equation}
If $P \, \RErr' x$, $L \, \RErr' p$  and $L P$ are all $\gg \hbar$ we have the approximate relation
\begin{equation*}
   \RErr' x \, \RErr' p \gtrsim \frac{\hbar}{2}
\end{equation*}
One expects this approximate form of the retrodictive error relationship to be valid in most 
situations of practical interest.  However, it is not always valid (see the Appendix for a counter example).

Starting from Eqs.~(\ref{eq:  ErrDisComRelA}) and~(\ref{eq:  ErrDisComRelB}) we can  derive in a similar manner
\begin{align}
\begin{split}
      \left( \RErr' x + \frac{\hbar}{P} \right) \, \left( \Dist' p + \frac{\hbar}{L} \right)
& \ge \frac{\hbar}{2} \left( 1 +\frac{2 \hbar}{LP} \right) \\
       \left( \RErr' p + \frac{\hbar}{L} \right) \, \left( \Dist' x + \frac{\hbar}{P} \right) 
& \ge  \frac{\hbar}{2} \left( 1 +\frac{2 \hbar}{LP} \right) 
\end{split}
\label{eq:  DistErrRelC}
\\
\intertext{and}
\begin{split}
       \PErr' x \, \left( \Dist' p + \frac{\hbar}{L} \right) & \ge \frac{\hbar}{2} \\
       \PErr' p \, \left( \Dist' x + \frac{\hbar}{P} \right) & \ge \frac{\hbar}{2}  
\end{split}
\label{eq:  DistErrRelD}
\end{align}
\section{Concluding Remarks}
The commonest method of describing the spread of a statistical distribution, in terms of the
variance---the method employed in this paper, in other words---is subject to certain 
limitations.  In recent years there has accordingly been some interest in devising
alternative approaches.  One approach is that involving parameter-based
uncertainty relationships~\cite{Hilgevoord,Braun}.  Another approach is
that involving entropic uncertainty relationships~\cite{BuschRetPre,ErrPrinciple,Ban,Buzek}.
It would be interesting to see if either of these approaches can be used to
develop the results obtained in this paper.

We should also remark, that in this paper we have made no use of the mathematical
theory based on the concept of a POVM, and an unsharp 
observable~\cite{Prugovecki,BuschUncert,deMuynck,BuschBook,Davies,Holevo,BuschRetPre,ErrPrinciple,Ban}.
There were certain advantages in proceeding in this way.  One advantage was, that it enabled us 
to circumvent the difficulties which have been identified by Uffink~\cite{Uffink,Appleby}.
Also, we share the view of Englert and W\'{o}dkiewicz~\cite{Wod2}, that 
the underlying intrinsic observables should be regarded as ``the heart of the matter.''
One of the advantages of the approach adopted here is, that it places 
the emphasis on these intrinsic observables, as opposed
to (in the words of Englert and W\'{o}dkiewicz) 
a ``mathematical representation of the statistical information gathered.''
Nevertheless, the theory of POVM's is clearly an important, and very powerful way of analysing 
simultaneous measurement processes. We certainly do not mean to set up the approach
taken in this paper as an \emph{alternative} to the approach based on POVM's.
We merely wish to stress the point made by
Englert and W\'{o}dkiewicz, that POVM's and unsharp observables should 
be regarded as secondary mathematical constructs, rather than as fundamental
physical concepts
which need to be posited from the outset.  We hope to return to this question
in a future publication, in which we will show how the concept
of an unsharp observable naturally emerges from the approach taken in this paper.
\section*{Appendix}
The purpose of this appendix is to explain why we defined the errors and disturbances by
taking the supremum over every normalised
initial system state, as in Eqs.~(\ref{eq:  MaxRMSRetErr}--\ref{eq:  MaxRMSDist}),
or a subset of them, as in Eqs.~(\ref{eq:  FinRangeRetErrs}--\ref{eq:  FinRangeDists}).  The reason is,
that there exist processes such that (for example)
\begin{equation*}
   \bmat{\psi \otimes \pt}{\Exi^2}{\psi \otimes  \pt}
   \bmat{\psi \otimes \pt}{\Epi^2}{\psi \otimes \pt}
=  0
\end{equation*}
for certain choices of initial system state $\ket{\psi}$ and initial 
apparatus state $\ket{\pt}$.  It is only when one takes the appropriate 
supremum that one gets the inequalities of
Eqs.~(\ref{eq:  RetErrRelB}) and~(\ref{eq:  DistErrRelB})
or Eqs.~(\ref{eq:  RetErrRelC}--\ref{eq:  DistErrRelD}).

Consider, for example, the measurement interaction described 
by the evolution operator
\begin{equation*}
  \hat{U} = \exp \left[- \frac{i \pi}{2 \hbar} \left( \xOp \PxOp - \MxOp \pOp \right) \right]
\end{equation*}
where $\PxOp$ is a momentum canonically conjugate to the pointer observable
$\MxOp$.  $\hat{U}$ is a rotation operator in $x p\Mx \Px$ space.  It
takes $\MxOp$ onto $\xOp$ and $\xOp$ onto $-\MxOp$:
\begin{equation*}
    \begin{pmatrix} \xfOp \\ \MxfOp \end{pmatrix}
= \hat{U}^{\dagger}
    \begin{pmatrix} \xOp\\\MxOp \end{pmatrix}
    \hat{U}
=   \begin{pmatrix}  \cos \frac{\pi}{2} & -\sin \frac{\pi}{2} \\
                      \sin \frac{\pi}{2} & \cos \frac{\pi}{2}
    \end{pmatrix}
    \begin{pmatrix}  \xOp \\ \MxOp \end{pmatrix}
=  \begin{pmatrix} - \MxOp \\ \xOp \end{pmatrix}
\end{equation*}
Similarly
\begin{equation*}
    \begin{pmatrix} \pfOp \\ \PxfOp \end{pmatrix}
=   \begin{pmatrix} - \PxOp \\ \pOp \end{pmatrix}
\end{equation*}
$\MpOp$ is unaffected by the interaction.  Referring back to the
definitions, Eqs.~(\ref{eq:  RetErrDef}-\ref{eq:  DistDef}), we deduce
\begin{equation*}
\begin{aligned}
  \Exi & = 0 & \hspace{0.5 in} \Exf & = \MxOp + \xOp  & \hspace{0.5 in} \Dx & =  - \MxOp- \xOp \\
  \Epi & = \MpOp - \pOp & \hspace{0.5 in} \Epf & = \MpOp + \PxOp & \hspace{0.5 in} \Dp & = - \PxOp - \pOp
\end{aligned}
\end{equation*}
Since $\MxfOp = \xiOp$ the process effects a perfectly accurate retrodiction of position, and this
is reflected in the fact that $\RErr x = 0$.  On the other hand the momentum pointer is unaffected
by the interaction:  $\MpfOp = \MpiOp$.  This means that the process is not really measuring the 
momentum at all.  We accordingly find $\RErr p = \infty$.  If we use the alternative definition
of Eq.~(\ref{eq:  FinRangeRetErrs}) then we find
\begin{equation*}
   \RErr' p \ge \frac{P}{2}
\end{equation*}
---which is again consistent with the fact, that so far as momentum is concerned, the process hardly
counts as a measurement.  Nevertheless, from the fact that
\begin{equation*}
   \bmat{\psi \otimes \pt}{\Epi^2}{\psi \otimes \pt} 
=  \left(\Delta \Mp \right)^2 + \left(\Delta p \right)^2 + \Bigl( \bmat{\pt}{\MpOp}{\pt} - \bmat{\psi}{\pOp}{\psi}\Bigr)^2
\end{equation*}
we see, that by appropriately choosing $\ket{\psi}$ and $\ket{\pt}$, $\left< \Epi^2 \right>$ can be
made arbitrarily small.  Moreover, the product $\left< \Exi^2 \right> \, \left< \Epi^2 \right>$ will
be zero whenever $\left< \Epi^2 \right>$ is finite. 

It is not surprising that $\left< \Epi^2 \right>$ is small for certain choices of  initial state.  
Consider, for example, the classical situation, where one has
a classical ammeter whose  needle is stuck at the 1 amp position.
Then the meter will, of course, give exactly the right reading if one uses it to measure a 
1 amp current.

\end{document}